\documentclass[acmsmall, nonacm]{acmart}
\usepackage{cleveref}
\usepackage{amsmath}
\theoremstyle{definition}
\newtheorem{definition}{Definition}
\usepackage{dsfont}
\usepackage{mathtools}
\newcommand\vldbdoi{XX.XX/XXX.XX}

\newcommand\vldbavailabilityurl{https://github.com/hereditary-eu/PrivEval}
\begin{document}
\title{A Review of Privacy Metrics for Privacy-Preserving Synthetic Data Generation}

%%
%% The "author" command and its associated commands are used to define the authors and their affiliations.
\author{Frederik Marinus Trudslev}
\orcid{0000-0001-5201-9326}
\affiliation{%
  \institution{Aalborg University}
  \city{Aalborg}
  \state{Denmark}
}
\email{fmtr@cs.aau.dk}

\author{Matteo Lissandrini}
\orcid{0000-0001-7922-5998}
\affiliation{%
  \institution{University of Verona}
  \city{Verona}
  \country{Italy}
}
\email{matteo.lissandrini@univr.it}

\author{Juan Manuel Rodriguez}
\orcid{0000-0002-1130-8065}
\affiliation{%
  \institution{Aalborg University}
  \city{Aalborg}
  \country{Denmark}
}
\email{jmro@cs.aau.dk}

\author{Martin Bøgsted}
\orcid{0000-0001-9192-1814}
\affiliation{%
  \institution{Aalborg University \& Aalborg University Hospital}
  \city{Aalborg}
  \country{Denmark}
}
\email{martin.boegsted@rn.dk}

\author{Daniele Dell'Aglio}
\orcid{0000-0003-4904-2511}
\affiliation{%
  \institution{Aalborg University}
  \city{Aalborg}
  \country{Denmark}
}
\email{dade@cs.aau.dk}

\maketitle

% \pagestyle{\vldbpagestyle}
% \begingroup\small\noindent\raggedright\textbf{PVLDB Reference Format:}\\
% \vldbauthors. \vldbtitle. PVLDB, \vldbvolume(\vldbissue): \vldbpages, \vldbyear.\\
% \href{https://doi.org/\vldbdoi}{doi:\vldbdoi}
% \endgroup
% \begingroup
% \renewcommand\thefootnote{}\footnote{\noindent
% This work is licensed under the Creative Commons BY-NC-ND 4.0 International License. Visit \url{https://creativecommons.org/licenses/by-nc-nd/4.0/} to view a copy of this license. For any use beyond those covered by this license, obtain permission by emailing \href{mailto:info@vldb.org}{info@vldb.org}. Copyright is held by the owner/author(s). Publication rights licensed to the VLDB Endowment. \\
% \raggedright Proceedings of the VLDB Endowment, Vol. \vldbvolume, No. \vldbissue\ %
% ISSN 2150-8097. \\
% \href{https://doi.org/\vldbdoi}{doi:\vldbdoi} \\
% }\addtocounter{footnote}{-1}\endgroup

\ifdefempty{\vldbavailabilityurl}{}{
\vspace{.3cm}
\begingroup\small\noindent\raggedright\textbf{Artifact Availability:}\\
The source code is available at \url{\vldbavailabilityurl}.
\endgroup
}

\section{Introduction}
Privacy Preserving Synthetic Data Generation (PP-SDG) has emerged to produce synthetic datasets from personal data while maintaining privacy and utility.
Differential privacy (DP) is the property of a PP-SDG mechanism that establishes how protected individuals are when sharing their sensitive data.
It is however difficult to interpret the privacy budget ($\varepsilon$) expressed by DP.
% To make the actual risk associated with the privacy loss more transparent, multiple libraries have been proposed to assess the privacy risk of the data using a multitude of privacy metrics (PMs).
% Most PMs available through these libraries are proposed in separate studies to assess newly introduced PP-SDG mechanisms.
To make the actual risk associated with the privacy budget more transparent, multiple privacy metrics (PMs) have been proposed to assess the privacy risk of the data.
These PMs are utilized in separate studies to assess newly introduced PP-SDG mechanisms \cite{patient-centric_synthetic_data_generation, healthGAN, yoon_anonymization_2020, medGAN}.
%Most PMs available through different libraries. 
Consequently, these PMs embody the same assumptions as the PP-SDG mechanism they were made to assess.
Therefore, a thorough definition of how these are calculated is necessary.

In this work, we present the assumptions and mathematical formulations of 17 distinct privacy metrics used in PrivEval \cite{priveval} to assess the risk of specific privacy attacks.

\section{Privacy Metrics}
To formally define the privacy metrics, we first define the setup in which this study is conducted. 

\begin{definition}[Dataset] 
   Let $A = \{a_1, a_2, \dots, a_m\}$ be a set of attributes, and let $t$ be a relation between attributes and their associated datatype, i.e. $t: A \rightarrow \{\mathbb{R}, \sum, \{0, 1\}
   \}$, meaning the data type of an attribute can be \textit{numerical}, \textit{categorical} or \textit{binary} respectively. A dataset is thereby defined as:
    \begin{equation}
        D \subset t(a_1) \times t(a_2) \times \dots \times t(a_m) 
    \end{equation}
    A tuple $d_j \in D$ holds a value for each attribute in $A$. 
\end{definition}

Let $A' \subseteq A$ be a subset of the attributes. Given $d_j\in D$, we indicate with $d_j[A']$ the projection of $d_j$ over the attributes in $A'$. 
It follows that $d_j=d_j[A]$.
Abusing the notation, we indicate with $D[A']$ the projection of all the tuples in $D$ over $A'$, i.e. $\{d_1[A'], \ldots, d_n[A']\}$, and with 
$d_j[a_i]$ the value of the attribute $a_i$ of the $j$-th instance.

Let $\mathcal{D}$ be the set of all datasets, and $Y\in\mathcal{D}$ be a dataset containing sensitive real data, and $Z\in\mathcal{D}$ be the synthetic dataset generated by the PP-SDG mechanism. 
We define a privacy metric as follows. 

\begin{definition}[Privacy Metric]\label{def:PM}
        A privacy metric $p$ quantifies the threat caused by the synthetisation mechanism: 
        \[
            p: \mathcal{D} \times \mathcal{D} 
            %\times \mathcal{D} 
            \rightarrow [0, 1]
        \]
        $p$ maps a real dataset and synthetic dataset 
        %and the auxiliary information 
        to a score in the range $[0,1]$, where $0$ indicates complete privacy and $1$ indicates no privacy. 
\end{definition}

Therefore, given a well-designed privacy metric $p_1$, one may expect that if $Z = Y$, then $p_1(Y, Z)$ should be $1$. 
Similarly, if $Y$ and $Z$ are totally independent from each other, one may expect that $p_1(Y, Z)$ should be very close to $0$.

\subsection{Simulation-based privacy metrics}
\begin{figure}[t!] 
  \centering
  \includegraphics[width=0.3\textwidth]{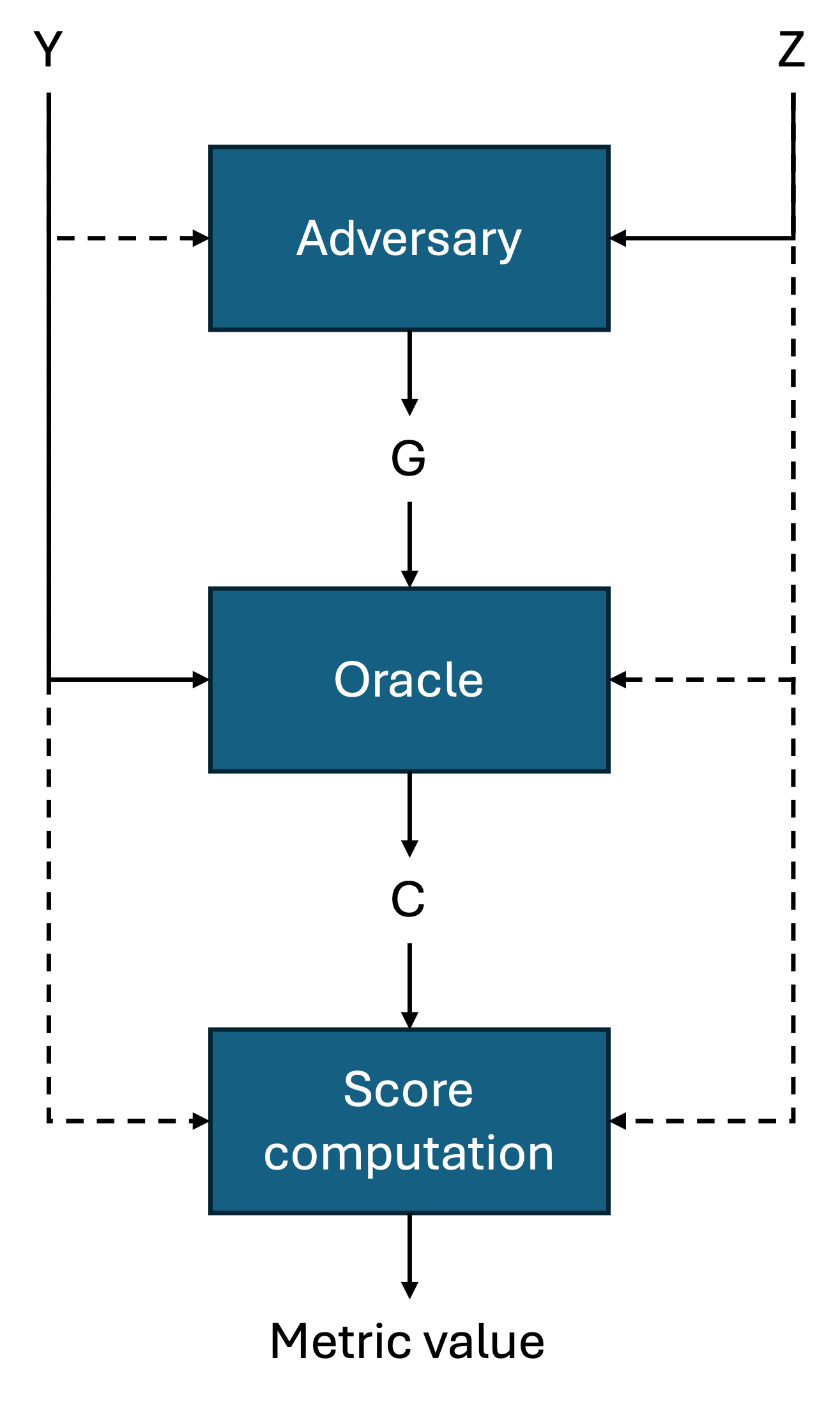}
  \caption{Metric computation in simulation-based privacy metrics.}
  \label{fig:attack} 
\end{figure}

The simulation-based privacy metrics simulate the presence of an adversary who runs a privacy attack. 
These metrics are usually computed as depicted in \autoref{fig:attack}.

The computation of these metrics start with an adversary who takes as input the synthetic dataset $Z$ and, optionally, some background knowledge as a portion of $Y$, and it captures a relation $G \subset Y \times Z$, denoting the \emph{guesses} of the attacker, defined as associations between elements in $Y$ and $Z$. 
The relation $G$ captures the attack.
This resulting pairs are then analysed by an oracle that determines the relation $C \subseteq R$ of correct guesses, and optionally the wrong and missing ones. 
By analysing the relations $C$ and $G$, the privacy score is computed.

\subsubsection{Categorical Zero CAP ($ZCAP$) \cite{CAP_metrics}}
$ZCAP$ measures the risk of an adversary inferring a categorical sensitive attribute ($a_S \in A$) using the key attributes $A_K \subseteq A \setminus \{a_S\}$, which can be continuous or categorical. 
$ZCAP$ assumes that the adversary knows the key attribute values of a real individual ($y[A_K]$) as well as all the attributes of the synthetic dataset $Z$.

$ZCAP$ simulates an attack where, for each real record $y\in Y$, the adversary finds matches between real and synthetic individuals based on the key attributes $A_K$, and then it checks how many of them share the sensitive values. 
Formally, we build the set of pairs of real and synthetic individuals sharing the same key attribute values:
\[G_{ZCAP}=\{(y,z) | y \in Y, z \in Z, y[A_K]=z[A_K]\},\]
and the subset $C$ of pairs that share both the key and sensitive attribute values: 
\[C_{ZCAP}=\{(y,z) | (y,z) \in G_{ZCAP}, y[A_S]=z[A_S]\}.\]

We can now define ZCAP as the ratio between the exact matches in $C_{ZCAP}$ and the total number of matches in $G_{ZCAP}$. 
Firstly, we define the metric for a real record $\hat{y}\in Y$ as:
\begin{equation}
    zcap(\hat{y}, Z) = 
    \frac
    {|\{z | (\hat{y},z) \in C_{ZCAP}\}|}
    {\max\{1,|\{z | (\hat{y},z) \in G_{ZCAP}\}|\}},
\end{equation}
where the denominator is the total number of matches, or $1$ when no matches exist. 
Secondly, we define $ZCAP$ for the whole dataset $Y$ as the average of the individual $ZCAP$ values:
\begin{equation}
    ZCAP(Y, Z) = \frac{1}{|Y|}\sum\limits_{y \in Y} zcap(y, Z).
\end{equation}

Therefore, $ZCAP$ varies between $0$ (no individual was reconstructed) to $1$ (all the individuals were reconstructed).

\subsubsection{Categorical Generalized CAP ($GCAP$) \cite{CAP_metrics}}
$GCAP$ extends $ZCAP$ by loosening the criterion of finding matching records by introducing a notion of similarity.
In $GCAP$, if a synthetic individual that has the same key values of the input individuals cannot be found, then the most similar synthetic individual(s) are used.
As $ZCAP$, $GCAP$ assumes that the adversary knows $y[A_K]$ and $Z$, and that sensitive attribute $a_S$ is categorical.

In $GCAP$, we build the matching set as:
\[G_{GCAP} = \{(y,z) | y \in Y, z = \arg\min_{z \in Z}d_H(y[A_K], z[A_K])\},\]
where $d_H$ indicates the Hamming distance. 
$G_{GCAP}$ contains all the real records in $Y$ and the closest synthetic record according to the vector space built on the key attributes. 
%The matching criteria in GCAP is the following: given an individual $y$, GCAP finds its nearest neighbour in the vector space induced by the key attributes $A_K$.
%Let $NN_H(y,Z)$ be the functions that retrieves the set of nearest neighbours of $y$ in $Z$ according to the Hamming distance, i.e.:
%\begin{equation}
    %NN_{H_K}(y, Z) = \arg\min\limits_{z \in Z} d_H(y[A_K], z[A_K])
%\end{equation}
%Where $d_H(y[A_K], z[A_K])$ is the Hamming distance between $y[A_K]$ and $z[A_K]$. 
The definition of $C_{GCAP}$ is analogous to the one of $C_{ZCAP}$:
\[C_{GCAP}=\{(y,z) | (y,z) \in G_{GCAP}, y[a_S]=z[a_S]\}.\]

Similarly to $ZCAP$, we can now first define $GCAP$ for a real individual $\hat{y}\in Y$:
\begin{equation}
    gcap(\hat{y}, Z) = 
    \frac
    {|\{z | (\hat{y},z) \in C_{GCAP}\}|}
    {|\{z | (\hat{y},z) \in G_{GCAP}\}|},
\end{equation}
and then $GCAP$ for the set of real records:
\begin{equation}
    GCAP(Y, Z) = \frac{1}{|Y|}\sum\limits_{y \in Y} gcap(y, Z).
\end{equation}

A value of $0$ indicates that no individual is correctly reconstructed, while $1$ indicates that the whole dataset was reconstructed.

\subsubsection{Attribute Inference Risk ($AIR$) \cite{multifaceted}}
$AIR$ is similar to $GCAP$, but it focuses on both discrete and continuous variables.
It finds matches of real records and similar synthetic records representing the key attributes in a vector space and using the Euclidean distance to measure similarity. 
To include categorical attributes, $AIR$ transforms them in numeric vectors through one hot encoding. 
The set of matching pairs is therefore defined as:
\[G_{AIR} = \{(y,z) | y \in Y, z = \arg\min_{z \in Z}d_E(y[A_K], z[A_K])\},\]
where $d_E$ is the Euclidean distance.

In $AIR$, the sensitive attribute $a_S$ can be either discrete or continuous.
When $a_S$ is discrete, the set of correct matches $C_{AIR}$ is defined as for $ZCAP$ and $GCAP$:
\[C_{AIR}=\{(y,z) | (y,z) \in G_{AIR}, y[a_S] = z[a_S]\}\]

When $a_S$ is continuous, $AIR$ introduces a notion of similarity also in the computation of the set of correct matches $C_{AIR}$.
Specifically, $AIR$ consider a correct match if two attributes' values differ at most by 10\%.
This leads to the following definition of $C_{AIR}$:
\[C_{AIR}=\{(y,z) | (y,z) \in G_{AIR}, |y[a_S] - z[a_S]| \leq z[a_S]*0.1)\}\]

$AIR$ quantifies uses the F1 score to quantify the success of the attack as a prediction task. 
To define the F1 score, we need to define the true positive, true negative and false negative matches. 
Here, $C_{AIR}$ is the set of true positive matches, $F_{AIR} = G_{AIR} \setminus C_{AIR}$ are the false positive matches, and the false negative match set is defined as the pairs including wrongly matched $y$ and for which it exists a synthetic record that satisfies the matching condition, i.e.:
\[M_{AIR}=\{(y,z) | \exists \bar{z} : (y,\bar{z}) \in F_{AIR}, z \in Z, y[a_S] = z[a_S]\},\]
when $a_S$ is a categorical variable, and:
\[
M_{AIR}= \{(y,z) | \exists \bar{z} : (y, \bar{z}) \in F_{AIR}, z \in Z, |y[a_S] - z[a_S]| \leq z[a_S]*0.1)\},    
\]
when $a_S$ is a continuous variable.

Let us define $TP=|C_{AIR}|$, $FP=|F_{AIR}|$ and $FN=|M_{AIR}|$, we can compute the F1 score as:
%If the predicted values $\bar{z}[A_S]$ partially or fully correspond to the actual values $y[A_S]$, the privacy is violated and therefore AIR increases.
%These violations are quantified through the F1-score. 
%We define the true positive ($TP$), false positive ($FP$) and false negative ($FN$) rate as:
\begin{equation}
    F_{1}(y[a], Z) = \frac{2 * \frac{TP}{TP+FP} * \frac{TP}{TP+FN}}{\frac{TP}{TP+FP} + \frac{TP}{TP+FN}}
\end{equation}

Finally, $AIR$ for the synthetic dataset is defined as:
\begin{equation}
    AIR(Y, Z) = \sum\limits_{y \in Y} w(y) * 
    %\sum_{a\in A_S} 
    F_{1}(y[a_s], Z),
\end{equation}
where $w(y)$ is a weight associated with a tuple $y$, such that: 
\begin{equation}
    w(y) = \frac{-P(y)  \phantom{i} log (P(y))}{H(Y)},
\end{equation}
where $P(y)$ is the probability of observing the set of attribute values of individual $y \in Y$.

To calculate the weight, we compute the entropy $H(Y)$ of the real dataset $Y$ as:
\begin{equation}
    H(Y) = -\sum\limits_{y \in Y} P(y)  \phantom{i} log (P(y))
\end{equation}

\begin{comment}
\begin{equation}
    \begin{aligned}
        TP &= 
        \begin{cases}
            |\{\bar{z} | \bar{z} \in NN_E(y, Z) \wedge y[a] = \bar{z}[a]\}| 
            & \text{if } a \in \sum \cup \{0,1\}  
            \\
            |\{\bar{z} | \bar{z} \in NN_E(y, Z) \wedge y[a]*0.9 < \bar{z}[a] < y[a]*1.1\}| 
            & \text{if } a \in \mathbb{R}
        \end{cases}
        \\
        FP &=
        \begin{cases}
            |\{\bar{z} | \bar{z} \in NN_E(y, Z) \wedge y[a] \neq \bar{z}[a]\}| 
            & \text{if } a \in \sum \cup \{0,1\}
            \\
            |\{\bar{z} | \bar{z} \in NN_E(y, Z) \wedge y[a]*0.9 \geq \bar{z}[a] \vee \bar{z}[a] \geq y[a]*1.1\}| 
            & \text{if } a \in \mathbb{R}
            
        \end{cases}
        \\
        FN &= 
        \begin{cases}
            |\{z | z \notin NN_E(y, Z) \wedge y[a] = z[a]\}| 
            & \text{if } a \in \sum \cup \{0,1\}  
            \\
            |\{z | z \notin NN_E(y, Z) \wedge y[a]*0.9 < z[a] < y[a]*1.1\}| 
            & \text{if } a \in \mathbb{R}
        \end{cases}
    \end{aligned}
\end{equation}

For continuous data to be considered correctly guessed, the sensitive attribute value has to be within a $10$ percent bound of the real value, where for categorical and binary values, the value has to be the same. 
The F1-score can be calculated as usual as:
\end{comment}

\subsubsection{Common Rows Proportion ($CRP$) \cite{synthcity}}
$CRP$ measures the risk of re-identification by computing the probability of a synthetic individual being the exact same as a real individual. 
$CRP$ assumes that an adversary has access to $Z$, and re-identify individuals from this through equality. 
In other words, it builds the matching set:
\[G_{CRP} = \{(z,z) | z \in Z\},\]
from which we can derive the set of correct matches as:
\[C_{CRP} = \{(z,z) | (z,z) \in G_{CRP}, z \in Y\}.\]
The probabilistic measure of $CRP$ is calculated as the ratio of correct matches, i.e.: 
\begin{equation}
    CRP(Y, Z) = \frac{|C_{CRP}|}{|Y| + 1\text{e-}8},
\end{equation}
where $1\text{e-}8$ is added to the denominator for numerical stability.
Thereby, a $CRP$ score of $0$ indicates that there are no identical real and synthetic individuals, and $CRP \approx 1$ indicates that all synthetic and real individuals are identical.

\subsection{Distance-based privacy metrics}
\begin{figure}[t!] 
  \centering
  \includegraphics[width=0.3\textwidth]{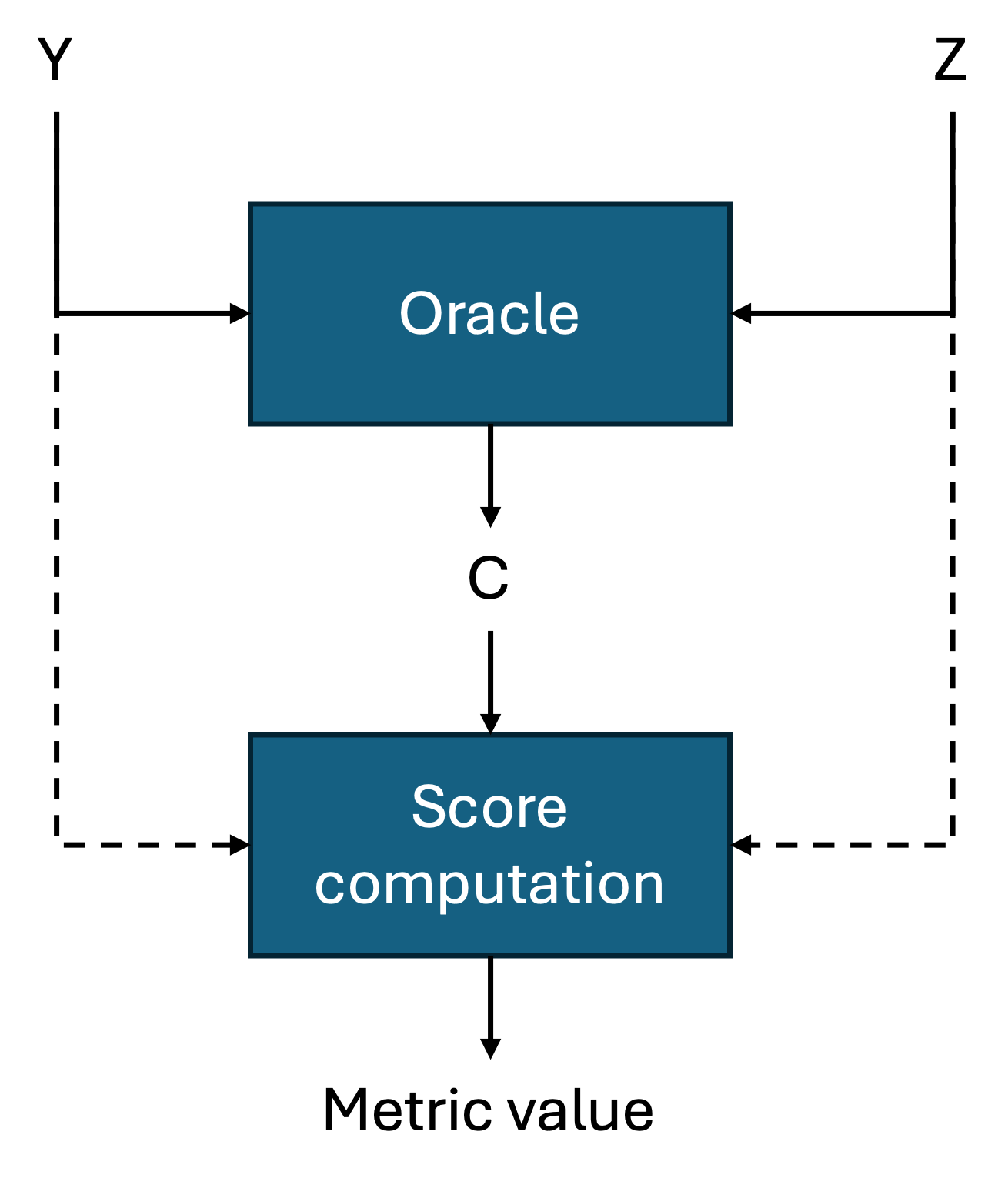}
  \caption{Metric computation in distance-based privacy metrics.}
  \label{fig:distance} 
\end{figure}

The distance-based privacy metrics do not simulate a privacy attack. 
Instead, they directly analyse the real and synthetic datasets $Y$ and $Z$, as depicted in \autoref{fig:distance}. 
The oracle analyses the two datasets and outputs a set of matches $C$, which for example includes the real records and their nearest neighbours. 
Next, the metric value is computed by analysing the set $C$, $Y$ and $Z$.

\subsubsection{Close Value Probability ($CVP$) \cite{synthcity}}
$CVP$ measures the probability that an adversary is able to distinct which real individual a synthetic is generated from, in order to.
%, where it is assumed that the adversary has access to $Z$, and wants to 
re-identify real individuals.
%from this.
$CVP$ measures the risk of re-identification as a probabilistic distance measure, where if the normalised distance from a real individual ($y$) to the nearest synthetic individual ($z$) is less than the threshold of $0.2$, the individual is deemed identifiable.
To calculate $CVP$, we first define the set of pairs where each real individual $y\in Y$ is matched to its closest synthetic individual $z\in Z$:

\begin{equation}\label{eq:ccvp}
C_{CVP} = \{(y,z) | y \in Y, z = \mathop{\arg\min}_{z \in Z}d_E(y, z)\},
\end{equation}
where $d_E$ is the Euclidean distance.
Differently from $ZCAP$, $GCAP$ and $AIR$, the distance is here considered in the vector space including all the attributes, and not only the key ones.

Let also 
\begin{align}
    \label{eq:dmax} d_{max} &= \max_{y\in Y,z\in Z}d_E(y,z) \\
    \label{eq:dmin} d_{min} &= \min_{y\in Y,z\in Z}d_E(y,z)
\end{align} 
be the maximum and minimum distance values between elements of $Y$ and $Z$.
$CVP$ is calculated as the average number of matches from $C_{CVP}$ whose normalised distance is lower than 0.2, i.e.:
\begin{equation}\label{normalised_nn}
    CVP(Y, Z) = \frac{|\{(y,z) | (y,z) \in C_{CVP}, \frac{d_E(y,z)-d_{min}}{d_{max}-d_{min}}\leq 0.2\}|
    }{|Y|}
\end{equation}
It means that an individual record is considered re-identifiable if it is \emph{close} to a synthetic record. 
A $CVP$ score of $1$ means that all real individuals are re-identifiable, while a score of $0$ means that none are re-identifiable.

\subsubsection{Distant Value Probability ($DVP$) \cite{synthcity}}
$DVP$ evaluates the distance-based separability of the real and synthetic data.
It is similar to $CVP$ as it considers the normalised distance between a real individual $y\in Y$ and its closest synthetic one.
In contrast, it checks whether the two individuals are \emph{far}.
Therefore, $DVP$ measures the inverse probability that a synthetic individual has a large distance to real individuals
%, while also assuming that an adversary has access to $Z$, 
and wants to re-identify real individuals from this.

The set of matching individuals is the same computed by $CVP$ in \autoref{eq:ccvp}, so  $C_{DVP} = C_{CVP}$.
We also consider the minimum and maximum distances between individuals $d_{max}$ and $d_{min}$ as defined in \autoref{eq:dmax} and \autoref{eq:dmin}.
The original definition of $DVP$ \cite{synthcity} defines it as:
\begin{equation}
    DVP(Y, Z) = \frac{|\{(y,z) | (y,z) \in C_{DVP}, \frac{d_E(y,z)-d_{min}}{d_{max}-d_{min}}\geq 0.8\}|
    }{|Y|},
\end{equation}
which indicates the probability of synthetic individuals being far away from the real individuals. 
This does not fit our definition of a privacy metric in \autoref{def:PM}, instead one can calculate it as:
\[
 DVP'(Y, Z) = 1 - DVP(Y, Z)
\]
A $DVP$ of $1$ thereby means there is no chance to have synthetic individuals far away from the real individuals, and a $DVP$ of $0$ means that all the synthetic individuals are far away from the real individuals.

\subsubsection{DetectionMLP ($D-MLP$) \cite{synthcity}}
$D-MLP$ assesses how well a simple binary classifier can distinguish between the synthetic and the real data. 
Specifically, it models a scenario where an adversary trains a shallow neural network to detect whether a given data point is real or synthetic.
We first create a set $D$ of labeled data, where to the real individuals in $Y$ is assigned the label $1$, and to the synthetic individuals in $Z$ is assigned the label $0$.
Next, a MLP is trained on $D$ using k-fold cross-validation, and 
the dataset $D$ is then separated in training and test sets ($D_{train}$ and $D_{test}$ resp.), and $D$-MLP score is the average AUC across all $k$ folds.

\begin{comment}
   
To calculate this, all real data points in $Y$ are labeled with $1$s, and all synthetic data points in $Z$ are labeled with $0$s:

\begin{equation}
\begin{aligned}
    L_Y &= [1, 1, \dots, 1], \quad |L_Y| = |Y|\\
    L_Z &= [0, 0, \dots, 0], \quad |L_Z| = |Z|\\
    D &= Y \cup Z\\
    L &= L_Y \cup L_Z
\end{aligned}
\end{equation}
\sloppy The combined dataset $D$ and label vector $L$ are then split into $k$ train/test folds:
$(D_{train_1}, D_{test_1}, \dots, D_{train_k}, D_{test_k})$ and $(L_{train_1}, L_{test_1}, \dots, L_{train_k}, L_{test_k})$.
For each fold, a shallow multi-layer perceptron (MLP) is trained to predict the label indicating whether a point is real or synthetic.
Let $f_{\theta_i}: \mathbb{R}^d \rightarrow [0,1]$ denote the MLP classifier trained on fold $i$:
\begin{equation}
f_{\theta_i} = \text{MLP}(D_{train_i}, L_{train_i})
\end{equation}
The classifier's output is a probability that a data point is real. The performance is evaluated using the Area Under the ROC Curve (AUC) for each fold, and the final $D$-MLP score is the average AUC across all $k$ folds:
\begin{equation}
DMLP(D, L) = \frac{1}{k} \sum\limits_{i=1}^k AUC(L_{test_i}, f_{\theta_i}(D_{test_i}))
\end{equation}
\end{comment}
A $D-MLP$ score close to $1$ indicates that real and synthetic data are easily distinguishable, implying low privacy.
Conversely, a score near $0.5$ suggests the classifier performs no better than random guessing, indicating high privacy.

\subsubsection{Authenticity ($Auth$) \cite{synthcity}}
Authenticity measures the risk of re-identification through connecting specific real individuals to synthetic individuals as a distance measure.
%, where it is assumed that the adversary has access to $Z$ and the real individual of interest $y$. 

To calculate $Auth$, we first define the set of triples where each real individual $y \in Y$ is matched to its closest real and synthetic individuals:
%$z \in Z$, as defined in \autoref{eq:gcvp}, i.e. $G_{Auth} = G_{CVP}$. 
%We then define the set of pairs where each real individual $y \in Y$ is matched to its closest real individual $y' \in Y \setminus \{y\}$:
%\begin{equation}
%    C_{Auth} = \{(y,y') | y \in Y, y' = \mathop{\arg \min}_{y' \in Y\setminus \{y\}}d_E(y, y')\}
%\end{equation}
\begin{equation}\label{eq:gauth}
C_{Auth}=\{(y, y', z)|y\in Y, \nonumber
y' = \mathop{\arg\min}_{y' \in Y\setminus \{y\}}d_E(y, y'),z = \mathop{\arg \min}_{z \in Z}d_E(y, z)
\}\nonumber
\end{equation}

\begin{comment}
For each real individual $y$, we compute the distance to its real match $d_{C}(y)$ and the distance to its synthetic match $d_{G}(y)$ as:

\begin{equation}\label{eq:neighbour_distances}
\begin{aligned} 
    d_{G}(y) &= d_E(y, z) \text{, where } (y, z) \in G_{Auth} 
    \\ 
    d_{C}(y) &= d_E(y, y') \text{, where } (y, y') \in C_{Auth} 
\end{aligned} 
\end{equation}
\end{comment}

The original definition of $Auth$ \cite{synthcity} is the average proportion of real individuals for whom the nearest real neighbour is closer than the nearest synthetic neighbour, meaning that a score closer to 1 means high privacy, such that:

\begin{equation} 
    Auth(Y, Z) = \frac{|\{(y,y',z) | (y,y',z)\in C_{Auth}, d_E(y,y') < d_E(y,z)\} |}{|Y|}
\end{equation}

This does not fit the definition in \autoref{def:PM}.
Therefore, to rescale $Auth$ to let it fit the definition, it can be subtracted from 1:
\[
    Auth'(Y, Z) = 1 - Auth(Y, Z),
\]

A high $Auth$ score (close to $1$) thereby means that most real individuals have synthetic individuals closer than the closest real individual (i.e. there are mostly non-authentic synthetic individuals in the data), while a $Auth$ score close to $0$ means that most synthetic individuals are authentic.

\subsubsection{Identifiability Score ($ID$) \cite{synthcity}}
The $ID$ score estimates the risk of re-identifying any real individual. %while only having access to $Z$.
Similarly to $Auth$, $ID$ estimates the probability that the distance to the closest synthetic individual is smaller than the distance from the closest real individual. 
The difference from $Auth$ is that $ID$ weights the influence of the attributes. 

Let $a \in A$ be an attribute in $A$. 
We use entropy to weight the attribute values of individuals:
\begin{equation}
    H(a,v) = - P(Y[a] = v) \log(P(Y[a] = v)),
\end{equation}
where $P(Y[a] = y[a])$ is the frequency of the value, i.e.
\[
P(Y[a] = v) = \frac{|\{y | y \in Y, y[a] = v\}|}{|Y|}.
\]

The entropy describes how diverse the values of an attribute are.
Here, a high entropy means more diverse attribute values, which deems the attribute less useful for re-identification, while a low entropy means low diversity in the value of the attribute, making it more useful for re-identification.

This, however, does not reflect the weighting that should be put on the attributes.
Therefore, a weight is introduced such that attributes with a high entropy have a low weight and attributes with a low entropy have a high weight:

\begin{equation} 
    w(a,v) = \frac{1}{H(a,v) + \varepsilon}
\end{equation}

The weighting is then used to construct a weighted version of the real and synthetic datasets $\hat{Y}$ and $\hat{Z}$, where for each element $y \in Y$ ($z \in Z$), there is an element $\hat{y} \in \hat{Y}$ ($\hat{z} \in \hat{Z}$) defined as follows:
\begin{equation} 
    \hat{y}[a] = \frac{y[a]}{w(a,y[a]) + \varepsilon}, \qquad \forall a \in A
\end{equation}

This representation captures how useful the attributes are in re-identification.

\begin{comment}
Using the weights, we define the transformation function $\phi: \mathbb{R}^m \times \mathbf{w} \rightarrow \mathbb{R}^m$ such that:

\begin{equation} 
    \hat{y} = \left(\frac{x_1}{w_1 + \varepsilon}, \frac{x_2}{w_2 + \varepsilon}, \ldots, \frac{x_m}{w_m + \varepsilon}\right) 
\end{equation}

\sloppy The transformation function is then used to construct a weighted version of the real ($\hat{Y} = \{\hat{y}_1, \dots, \hat{y}_{|Y|}\}$ and synthetic ($\hat{Z} = \{\hat{z}_1, \dots, \hat{z}_{|Z|}\}$ datasets such that:
\begin{equation} 
\begin{aligned}
    \hat{Y} &= {\phi(y_i, \mathbf{w}) : y_i \in Y} = {\hat{y}_1, \hat{y}_2, \ldots, \hat{y}_{|Y|}} \\
    \hat{Z} &= {\phi(z_k, \mathbf{w}) : z_k \in Z} = {\hat{z}_1, \hat{z}_2, \ldots, \hat{z}_{|Z|}} 
\end{aligned}
\end{equation}
\end{comment}

We now define the set of triples of elements in $\hat{Y}$ and their nearest real and synthetic neighbours as: 

\[
    C_{ID} = \{(\hat{y}, \hat{y}', \hat{z}) | \hat{y} \in \hat{Y}, \nonumber 
    \hat{y}' = \mathop{\arg \min}_{y' \in \hat{Y}\setminus\{\hat{y}\}} d_E(\hat{y}, \hat{y}'),\nonumber
    \hat{z} = \mathop{\arg \min}_{z \in \hat{Z}} d_E(\hat{y}, \hat{z})\}\nonumber
\]

Here, the weighting influences the attributes' contribution to distance such that a low weight means a low contribution to the distance measure and vice versa. 
\begin{comment}
    We then modify $C_{Auth}$ to be the set of pairs, where each weighted real individual $y \in Y$ is matched to its closest weighted real individual $y' \in Y \setminus \{y\}$:
\begin{equation}
    C_{ID} = {(y_i, y_j) : y_i \in \hat{Y}, y_j = \arg\min_{y' \in \hat{Y} \setminus \{\hat{y}_i\}} d_E(\hat{y}_i, \hat{y}')}
\end{equation}

For each real individual $y_i$, we modify the comparison in equation \autoref{eq:neighbour_distances} such that we compute the weighted distance to its real match $d_{C}(y_i)$ and the distance to its synthetic match $d_{G}(y_i)$:

\begin{equation}
\begin{aligned} 
    d_{G}(y_i) &= d_E(\hat{y}_i, \hat{z}) \text{, where } (y_i, z) \in G_{ID}
    \\ 
    d_{C}(y_i) &= d_E(\hat{y}_i, \hat{y}) \text{, where } (y_i, y) \in C_{ID}
\end{aligned} 
\end{equation}
\end{comment}

The $ID$ score can then be calculated as the proportion of real individuals with matching weighted synthetic individuals closer than the matching weighted real individual, and can be calculated as:

\begin{equation}
    ID(Y, Z) = \frac{|\{(\hat{y}, \hat{y}', \hat{z}) | (\hat{y}, \hat{y}', \hat{z}) \in C_{ID}, d_E(\hat{y},\hat{z}) < d_E(\hat{y},\hat{y}')\}|}{|\hat{Y}|}
\end{equation}

Thereby, if a real individual has a matching real individual further closer than the matching synthetic individual, the individual is not at risk of re-identification, while if the opposite holds, the individual is at risk.
An $ID$ score near $1$ therefore means that most individuals are re-identifiable, while a score near $0$ means that most individuals are not re-identifiable.

\subsubsection{Nearest Synthetic Neighbour Distance ($NSND$) \cite{synthcity}}
$NSND$ quantifies the extent to which a synthetic individual can be traced back to the real individual from whom it was derived.
$NSND$ measures the re-identification risk as a normalised average of the distance of each real individual $y$ and its nearest synthetic individual $z$. 
The $NSND$ score is a measure of closeness, and measures the overall re-identification risk of the synthetic dataset.
%, where it is assumed that the adversary has access to $Z$ and wants to infer specific individuals.
We first compute the match set as in \autoref{eq:ccvp}, i.e. $C_{NSND} = C_{CVP}$.
Let also $d_{max}$ and $d_{min}$ be the maximum and minimum distances between a real and a synthetic individual, as defined in \autoref{eq:dmax} and \autoref{eq:dmin} respectively.
We compute $NSND$ as:
\begin{equation}
    NSND(Y, Z) = \frac{1}{|Y|} \sum_{(y,z)\in C_{NSND}} \frac{d_E(y,z)-d_{min}}{d_{max}-d_{min}}
\end{equation}

\begin{comment}
    
To normalise the nearest neighbour distances, we need to calculate the nearest neighbour distance between all real and synthetic individuals. This can be calculate as:

\begin{equation}\label{eq:dists}
    dists_E(Y, Z) = \{x_i | x_i = D_E(y_i, NN_E(y_i, Z)), \forall i \in [1 \dots n]\}
\end{equation}

where $NN_E(\cdot)$ is the nearest neighbour algorithm from equation \ref{normalised_nn}, and $D_E(\cdot)$ is the euclidean distance.
Using this, the nearest neighbour distances and min-max normalised, and the mean of these can be calculated as such:

\begin{equation}
    NSND(Y, Z) = \frac{\sum\limits_{x \in dists(Y, Z)} \frac{x - min(dists(Y, Z))}{max(dists(Y, Z)) - min(dists(Y, Z))}}{n}
\end{equation}
\end{comment}

If all the distances contributing to NSND are small (close to the minimum distance), the normalised distances will all be close to $0$, resulting in a low $NSND$.
Conversely, if the distances are disperse and include large distances (closer to the maximum), the normalised distances will increase, increasing the $NSND$.

\subsubsection{Nearest neighbour Distance Ratio ($NNDR$) \cite{patient-centric_synthetic_data_generation}}
$NNDR$ estimates the risk of re-identifying any real individual as a distance measure. %, while only having access to $Z$.
$NNDR$ initially maps real and synthetic individuals to a lower-dimensional space, and then compares distances between records.
Let 
\begin{equation}\label{eq:dimred}
\Phi^k:\mathbb{R}^{|A|}\rightarrow\mathbb{R}^{k}
\end{equation}
be a mapping that maps a space with $|A|$ dimensions to a $k$-dimensional Euclidean space% ($k=2$ in \textcolor{red}{CITE PRIVEVAL})
, e.g. by using PCA, MCA or FAMD.
Let also the set of matching elements $C_{NNDR}$ be the set of matches between synthetic individuals $z\in Z$ and their top-2 closest neighbours from $Y$ in the $k$-dimensional space, i.e.:
\[
    C_{NNDR} = \{(y',y'',z) z \in Z, \nonumber
    y' = \mathop{\arg\min}_{y' \in Y}d_E(\Phi^k(y'), \Phi^k(z))\},
    y'' = \mathop{\arg\min}_{y'' \in Y \setminus\{y'\}}d_E(\Phi^k(y''), \Phi^k(z))\} \nonumber
\]
%and let $G'_{NNDR}$ be the set of matches between synthetic individuals $z\in Z$ and their second closest neighbours from $Y$ in the $k$-dimensional space, i.e.:
%\begin{equation}
%    G'_{NNDR} = \{(y,z) | z \in Z, y = \mathop{\arg \min}_{y \in Y\setminus\{\bar{y}\}:(\bar{y},z)\in G_{NNDR}}d_E(\Phi^k(y), \Phi^k(z))\}.
%\end{equation}
We define $C'_{NNDR}$ as the set of pairs in $C_{NNDR}$ where the distance between the synthetic record and its closest real record is 0:
\[
    C'_{NNDR} = \{(y',y'',z) | (y',y'',z) \in C_{NNDR},  
    d_E(\Phi^k(y'), \Phi^k(z))=0\}
\]
We can now define NNDR.
Given a synthetic individual $z$, the associated score is $1$ if the distance to its closest neighbour in $Y$ is $0$, and the ratio between the distance to its closest real neighbour and the distance to the second closest real neighbour otherwise.
NDRR is the average of the scores, i.e.:
\begin{equation}
    NNDR(Y, Z) = \frac{\left(|C'_{NNDR}|+ \sum\limits_{(y',y'',z)\in C_{NNDR}\setminus C'_{NNDR}}\frac{d_E(y',z)}{d_E(y'',z)}\right)}{|Z|}
\end{equation}

\begin{comment}
, such that $\Phi^k: \mathcal{D} \rightarrow \mathbb{R}^k$. This mapping can therefore map $Y$ and $Z$ to the same $k$-dimensional space, such that

\begin{equation}\label{eq:pca}
\begin{aligned}
    \Phi^k(Y) &= Y' = \{y'_1, y'_2, \dots, y'_n\}, \\
    \Phi^k(Z) &= Z' = \{z'_1, z'_2, \dots, z'_l\}
\end{aligned}
\end{equation}

are the $k$-dimensional representations of individuals in the real $Y'$ and synthetic $Z'$ datasets. 
The score for a synthetic individual is then calculated in terms of distance, where if the distance to a real individual is $0$, the score is $1$.
Otherwise, if the distance is more than $0$, then the score for that synthetic individual is the distance to the closest real neighbour divided by the distance to the second closest real neighbour.
The score for a synthetic individual is thereby:

\begin{equation}
    r(Y', z') = 
    \begin{cases}
        1 & \text{if } D_E(z', NN_E(z', Y')) = 0, \\
        \frac{D_E(z', NN_E(z', Y'))}{D_E(z', 2NN_E(z', Y'))} & \text{otherwise}
    \end{cases}
\end{equation}

where $NN_E(\cdot)$ and $2NN_E(\cdot)$ are the same as the nearest neighbour algorithms from equations \ref{normalised_nn} and \ref{eq:2NNE}, respectively.
The $NNDR$ can then be calculated as the average $r$ score over all synthetic individuals as:

\begin{equation}
    NNDR(\Phi^k(Y), \Phi^k(Z)) = \frac{\sum\limits_{j=1}^l r(Y', z'_j)}{l}
\end{equation}
\end{comment}

Thereby, if $NNDR = 1$ means that all real individuals are re-identifiable, while if $NNDR \approx 0$, none are re-identifiable.

\subsubsection{Distance to Closest Record ($DCR$) \cite{patient-centric_synthetic_data_generation}}
$DCR$ estimates the risk of re-identification
%, while having access to $Z$, 
as an overall score derived from a measure of distance.
The first step is to compute the set of pairs of real individuals and their nearest synthetic individuals. 
Similarly to $NNDR$, the closeness is computed a in a smaller vector space, using the the same mapping $\Phi^k$ as $NNDR$ (\autoref{eq:dimred}), which maps real and synthetic individuals to a lower dimensional space.
Therefore, we define $C_{DCR}$ as:
\[
    C_{DCR} = \{(y,z) | y \in Y, z = \mathop{\arg\min}_{y \in Z}d_E(\Phi^k(y), \Phi^k(z))\}
\]

Next, $DCR$ computes the average distance of the pairs in $C_{DCR}$:
\begin{equation}
DCR(Y, Z) = \frac{1}{|Y|} \sum_{(y,z)\in C_{DCR}}d_E(\Phi^k(y),\Phi^k(z))
\end{equation}

The original definition of $DCR$ \cite{patient-centric_synthetic_data_generation} can assume any value, according to how the individuals are mapped in the vector space.
One can rescale $DCR$ to let it fit in a $[0,1]$ interval, for examples as:
\begin{equation}
    DCR'(Y, Z) = 1 - \sigma \left( log\left(DCR(Y,Z) \right)\right),
\end{equation}
where $\sigma$ is the sigmoid function used to fit the score in the desired range. 
Thereby, a $DCR'$ score close to $0$ means low re-identification risk, while a score close to $1$ means high re-identification risk. 

\subsubsection{Median Distance to Closest Record ($MDCR$) \cite{syntheval}}
$MDCR$ estimates the risk of re-identification % while having access to $Z$.
%$MDCR$ is 
by calculating the ratio of the median distance between the real and synthetic records, and the median among the real records:
\begin{equation}
    MDCR(Y, Z) = \frac{med\left(\{d_E(y, z) | y \in Y, z=\mathop{\arg\min}\limits_{z \in Z} d_E(y, z) \}\right)}
    {med\left(\{d_E(y, y') | y \in Y, y'=\mathop{\arg\min}\limits_{y' \in Y\setminus \{y\}} d_E(y, y') \}\right)},
\end{equation}
where $med(\cdot)$ is the median value of a set.

As for $DCR$, $MDCR$ can assume any value, depending on the vector space where distances are computed. 
We can scale the metric to fit in the $[0,1]$ range:
\begin{equation}
    MDCR'(Y, Z) = \sigma \left(MDCR(Y, Z) \right)
\end{equation}
where $\sigma$ is the sigmoid function.
A $MDCR'$ score close to $0$ means low re-identification risk, while a score close to $1$ means high re-identification risk. 

\subsubsection{Nearest Neighbour Adversarial Accuracy ($NNAA$) \cite{syntheval}}
$NNAA$ estimates the risk of determining whether an individual contributed their data to the real dataset 
while assuming that %the adversary has access to $Z$ and the person of interest, and 
the sets of real and synthetic individuals have the same size, i.e. $|Y|=|Z|$.
The score is calculated as two probabilities: (i) the probability that real data points are closest to a synthetic record than another real one, (ii) the probability that a synthetic record is nearer to a synthetic individual than a real one. 
To estimate the probabilities, we build the sets $C^Y_{NNAA}$ and $C^Z_{NNAA}$. 
\sloppy The former is defined as in \autoref{eq:gauth} (i.e. $C^Y_{NNAA}=C_{Auth}$), while the latter is defined as:
\[
    C^Z_{NNAA} = \{(y,z',z)|z\in Z, \nonumber
    y = \mathop{\arg \min}_{y \in Y}d_E(y, z), 
    z' = \mathop{\arg \min}_{z' \in Z \setminus \{z\}} d_E(z, z')\}\nonumber
\]

$NNAA$ is then computed as:
\begin{align}
NNAA(Y,Z) = \nonumber    
&\frac{|\{(y,y',z)|(y,y',z)\in C^Y_{NNAA}, d_E(y,z) > d_E(y,y')\}}{|C^Y_{NNAA}|}| +\\
&\frac{|\{(y,z',z)|(y,y',z)\in C^Z_{NNAA}, d_E(y,z) > d_E(z,z')\}|}{|C^Z_{NNAA}|} \nonumber
\end{align}

The score is then subtracted from 1 to match the direction of our definition of a privacy metric, and can be calculated as:
\begin{equation}
    NNAA'(Y,Z) = 1 - NNAA(Y,Z)       
\end{equation}
\begin{comment}
    
\begin{equation}
\begin{aligned}
    NNAA(Y, Z) = 1 - \frac{\sum\limits_{i=1}^n\splitdfrac{ \mathbf{1}[
    D_{E}(y_i, NN_E(y_i,Z))
    > 
    D_{E}(y_i, NN_E(y_i,Y))]}{
    +
    \mathbf{1}[
    D_{E}(z_i, NN_E(z_i,Y))
    > 
    D_{E}(z_i, NN_E(z_i, Z))]}
    }
    {2n}
\end{aligned}
\end{equation}

Here, $D_{E}(d, NN_E(d, D)$ denotes the Euclidean distance between an individual $d$ and its' nearest neighbour in $D$, $NN_E(\cdot, \cdot)$ is the Euclidean nearest neighbour defined in equation \autoref{normalised_nn} and $D_E(\cdot, \cdot)$ is the Euclidean distance.
\end{comment}
Thereby, a score of $0.5$ means that an adversary has equal chance of guessing whether any individuals contributed to the data or not, while if $NNAA = 1$ an adversary can determine the contribution for all individuals in the dataset.

\subsubsection{Membership Inference Risk ($MIR$) \cite{syntheval}}
$MIR$ estimates the risk of determining whether an individual contributed their data to the real dataset. 
%while assuming that the adversary has access to $Z$ and the person of interest.
$MIR$ is similar to $D-MLP$: it trains a classifier to distinguish the real and synthetic individuals, using $Y$ and $Z$.
It first creates a dataset $D$ with the individuals in $Y$ and $Z$, assigning them $1$ and $0$ as labels, respectively.
After splitting $D$ in training and test set, $MIR$ trains a LigtGBM classifier on the former, and uses the latter to compute the score as the recall (i.e. the ratio between the correctly labeled real individuals and the real individuals).
\begin{comment}
    
Similar to the calculation of $D-MLP$, all real data points in $Y$ are labeled with $1$s, and all synthetic data points in $Z$ are labeled with $0$s:

\begin{equation}
\begin{aligned}
    L_Y &= [1, 1, \dots, 1], \quad |L_Y| = |Y|\\
    L_Z &= [0, 0, \dots, 0], \quad |L_Z| = |Z|\\
    D &= Y \cup Z\\
    L &= L_Y \cup L_Z
\end{aligned}
\end{equation}

The combined dataset $D$ and label vector $L$ are then split into train/test stes ($D_{train}, D_{test}$) and ($L_{train}, L_{test}$).

Let $f_\theta: \mathbb{R}^d \rightarrow [0,1]$ be a classifier trained on $D_{train}$ using the LightGBM algorithm such that $f_\theta = LightGBM(D_{train})$
The prediction performed by the classifier is then:
\begin{equation}
    \hat{l}_i = 
    \begin{cases}
        1 & \text{if}\ f_\theta(x_j) \geq 0.5
        \\
        0 & \text{otherwise}
    \end{cases}
\end{equation}

The $MIR$ is then calculated as the recall of the classification task:

\begin{equation}
    MIR(D) = \frac{\sum\limits_{(x_j, l_j) \in D_{\text{test}}} \mathbf{1}\left[f_{\theta}(x_j) \geq 0.5 \,\land\, l_j = 1\right]}{\sum\limits_{(x_j, l_j) \in D_{\text{test}}} \mathbf{1}\left[l_j = 1\right]}
\end{equation}
\end{comment}
Thereby, an $MIR$ score close to $1$ suggests that the classifier can reliably identify real individuals, even from unseen samples, while an $MIR$ score close to $0$ suggests that the classifier cannot identify the individuals from the synthetic data. 

\subsubsection{Hidden Rate ($HiddR$) \cite{patient-centric_synthetic_data_generation}}
$HiddR$ estimates the risk of determining whether an individual contributed their data to the real dataset.
%while assuming that the adversary has access to $Z$ and the person of interest.
$HiddR$ assumes that each synthetic individual $z$ is generated from a real individual $y$. 

$HiddR$ checks, for each synthetic individual $z$, whether its nearest real neigbour is the same from which it was generated. 

Let $g$ be the relation that maps a real individual $y$ to the synthetic individual $z$ generated from itself . 
Let also $\Phi^k$ be a function that maps an individual to a smaller vector space, as defined in \autoref{eq:dimred}. 
We define the set of triples composed by the real individuals, and their nearest neigbhours:
\[
C_{HiddR}=\{(y,z)|y\in Y, z = \arg\min_{z\in Z}d_M(\Phi^k(y),\Phi^k(z))\},
\]
where $d_M$ denotes the Minkowski distance.
We define $HiddR$ as follows:
\begin{equation}
    HiddR(Y,Z) = \frac{|\{(y,z)|(y,z) \in C_{HiddR}, z = g(y) \}|}{|Z|}
\end{equation}

\begin{comment}
synthetic uses the same mapping as $NNAA$, $DCR$ and $NNDR$, where real and synthetic individuals are initially mapped to a lower dimensional space, and then distances are computed, as elaborated in equation \autoref{eq:pca}.
The metric assumes that synthetic samples are generated in a one-to-one manner and preserve the order, i.e., synthetic point $j$ is meant to correspond to real point $j$, and can be calculated as:

\begin{equation}
    HiddR(\Phi^k(Y), \Phi^k(Z)) = \frac{\sum\limits_{i=1}^n 1[NN_M(\Phi^k(y_i), \Phi^k(Z)) = \Phi^k(y_i)]}{n}
\end{equation}

Where $NN_M$ is the nearest neighbour algorithm in equation \autoref{normalised_nn} using Manhattan distance.
\end{comment}

If $HiddR$ is close to $1$, it suggests that synthetic data points strongly resemble their real counterparts, and thereby, there is a high risk of determining individual contributions, while if $HiddR$ is close to $0$, i suggests that the synthetic data is more diverse and decoupled from the real data.

\subsubsection{Hitting Rate ($HitR$) \cite{syntheval}}
$HitR$ estimates the risk of determining whether an individual contributed their data to the real dataset.
%while assuming that the adversary has access to $Z$ and the person of interest.
The metric counts how many synthetic and real individuals are \emph{close}, i.e. they have the same value for categorical attributes, and similar values for continuous variables. 

Let $A_c \subseteq A$ be the set of categorical attributes, and $A_r \subseteq A$ be the set of continuous attributes. 
We define the set of correct matches $C_{HitR}$ as:
\[C_{HitR} = \{(y,z)|y \in Y, z \in Z, y[A_c] = z[A_c], y[A_r] \sim z[A_r]\},\]
where $y[A_r] \sim z[A_r]$ indicates that the continuous values are within an attribute-dependent threshold $h(a)$, i.e. $\forall a \in A_r, |y[a]-z[a|] \leq h(a)$, defined as:
\begin{equation}
h(a) = \frac{\max\limits_{y\in Y} y[a] - \max\limits_{y\in Y} y[a]}{30}
\end{equation}

We can now define $HitR$ as:
\begin{equation}
HitR(Y, Z) = \frac{|C_{HitR}|}{|Y|}.
\end{equation}

%where $thr(y, z, A)$ determines whether it is within the threshold, and $h(a)$ is the threshold value which continuous values must be within for approximate equality, and is defined as:

Thereby, a score close to $1$ means that many individuals are at high risk of inference, while a score close to $0$ means that all individuals are distinct from the synthetic data.

\begin{acks}
This work is partially supported by the HEREDITARY Project, as a part of the European Union’s Horizon Europe research and innovation programme under Grant Agreement No GA 101137074.
\end{acks}

%%%%%%%%% no space for this now
%%%%% we add later

% \begin{acks}
%  %This work was supported by the [...] Research Fund of [...] (Number [...]). Additional funding was provided by [...] and [...]. We also thank [...] for contributing [...].
%  This work is partially supported by the HEREDITARY Project, as a part of the European Union’s Horizon Europe research and innovation programme under Grant Agreement No GA 101137074.
% \end{acks}

%\clearpage

\bibliographystyle{ACM-Reference-Format}
\bibliography{bibliography}

\end{document}